\documentclass[12pt,thmsa]{article}
\usepackage{sw20lart}

\input tcilatex
\QQQ{Language}{
American English
}

\begin{document}

\author{I. Radinschi\thanks{%
iradinsc@phys.tuiasi.ro} \and Department of Physics, ``Gh. Asachi''
Technical University, \and Iasi, 6600, Romania}
\title{Energy Distribution of a Charged Regular Black Hole }
\date{}
\maketitle

\begin{abstract}
We calculate the energy distribution of a charged regular black hole by
using the energy-momentum complexes of Einstein and M\o ller.


\end{abstract}

\section{INTRODUCTION}

One of the most interesting and intricate problem of relativity is the
energy-momentum localization. The different attempts at constructing an
energy-momentum density don't lead to a generally accepted expression.
However, there are various energy-momentum complexes including those of
Einstein [1], Tolman [2], Landau and Lifshitz [3], Papapetrou [4], Bergmann
[5], Weinberg [6] and M\o ller [7]. Cooperstock [8] gave his opinion that
the energy and momentum are confined to the regions of non-vanishing
energy-momentum tensor of the matter and all non-gravitational fields.
Although, the energy-momentum complexes are coordinate dependent they can
give a reasonable result if calculations are carried out in Cartesian
coordinates. Some interesting results obtained recently lead to the
conclusion that different energy-momentum complexes give the same energy
distribution for a given space-time [9]-[16].

We calculate the energy distribution of a charged regular black hole by
using the energy-momentum complexes of Einstein and M\o ller. We use the
geometrized units $(G=1,c=1)$ and follow the convention that Latin indices
run from $0$ to $3$.

\section{Energy in the Einstein prescription}

The Reissner-Nordstr\"{o}m (RN) metric is the only static and asymptotically
flat solution of the Einstein-Maxwell equations and it represents an
electrically charged black hole. The metric is given by

\begin{equation}
ds^2=A(r)dt^2-B(r)dr^2-r^2(d\theta ^2+\sin ^2\theta d\varphi ^2)  \label{1.1}
\end{equation}

where

\begin{equation}
A(r)=B^{-1}(r)=1-\frac{2M}r+\frac{q^2}{r^2}  \label{1.2}
\end{equation}
and $q$ and $M$ are the electric charge and, respectively, the mass of the
black hole.

A solution to the coupled system of the Einstein field and equations of the
nonlinear electrodynamics was recently given by E. Ay\'{o}n-Beato and
A.Garcia (ABG) [17]. This solution represents a regular black hole with mass 
$M$ and electric charge $q$ and avoids thus the singularity problem. Also,
the metric asymptotically behaves as the Reissner-Nordstr\"{o}m solution.
The usual singularity of the RN solution, at $r=0$, has been smoothed out
and now it simply corresponds to the origin of the spherical coordinates.
The line element is given by (1) with

\begin{equation}
A(r)=B^{-1}(r)=1-\frac{2M}r(1-\tanh (\frac{q^2}{2Mr})).  \label{1.3}
\end{equation}
If the electric charge vanishes we reach the Schwarzschild solution. At
large distances (3) resembles to the Reissner-Nordstr\"{o}m solution and can
be written

\begin{equation}
A(r)=B^{-1}(r)=1-\frac{2M}r+\frac{q^2}{r^2}-\frac{q^6}{12M^2r^4}+O(\frac
1{r^6}).  \label{1.4}
\end{equation}

We obtain the energy distribution associated with the solution given by (1)
and (3) in the Einstein and M\o ller prescriptions.

The Einstein energy-momentum complex [1] is given by

\begin{equation}
\Theta _i^{\;\,k}=\frac 1{16\pi }H_{i\;\;,l}^{\;kl},  \label{1.5}
\end{equation}

where

\begin{equation}
H_i^{\;kl}=-H_i^{\;lk}=\frac{g_{in}}{\sqrt{-g}}[-g[g^{kn}g^{lm}-g^{\ln
}g^{km})]_{,l}.  \label{1.6}
\end{equation}

$\Theta _0^{\;\,0}$ and $\Theta _\alpha ^{\;\,0}$ are the energy and,
respectively, the momentum components.

The Einstein energy-momentum complex satisfies the local conservation laws

\begin{equation}
\frac{\partial \Theta _i^{\;\,k}}{\partial x^k}=0.  \label{1.7}
\end{equation}

The energy and momentum in the Einstein prescription are given by

\begin{equation}
P_i=\int \hskip-7pt\int \hskip-7pt\int \Theta _i^{\;0}dx^1dx^2dx^3.
\label{1.8}
\end{equation}

Using the Gauss theorem we obtain

\begin{equation}
P_i={\frac 1{8\pi }}\int \hskip-7pt\int H_i^{\;0\alpha }n_\alpha dS,
\label{1.9}
\end{equation}
where $n_\alpha =\left( {\frac xr},{\frac yr},{\frac zr}\right) $ are the
components of a normal vector over an infinitesimal surface element $%
dS=r^2\sin \theta d\theta d\varphi $.

The required nonvanishing components of $H_i^{\;kl}$ for the line element
given by (1) and (3) are

\begin{equation}
\begin{tabular}{c}
$H_0^{\;01}=-\frac{4Mx(-1+\tanh (\frac{q^2}{2Mr}))}{r^3},$ \\ 
$H_0^{\;02}=-\frac{4My(-1+\tanh (\frac{q^2}{2Mr}))}{r^3},$ \\ 
$H_0^{\;03}=-\frac{4Mz(-1+\tanh (\frac{q^2}{2Mr}))}{r^3}.$%
\end{tabular}
\label{1.10}
\end{equation}

Using the above expressions in (9) we obtain

\begin{equation}
E(r)=M(1-\tanh (\frac{q^2}{2Mr})).  \label{1.11}
\end{equation}

From (11) it results that if $q=0$ we have the energy of a Schwarzschild
black hole.

After some calculations, we get the energy distribution of the ABG black hole

\begin{equation}
E(r)=M-\frac{q^2}{2r}+\frac{q^6}{24M^2r^3}-\frac{q^{10}}{240M^4r^5}+O(\frac
1{r^6}).  \label{1.12}
\end{equation}

Also, (12) can be written

\begin{equation}
E(r)=E_{RN}(r)+\frac{q^6}{24M^2r^3}-\frac{q^{10}}{240M^4r^5}+O(\frac 1{r^6}).
\label{1.13}
\end{equation}
\[
\]

\section{Energy in the M\o ller prescription}

The M\o ller energy-momentum complex [7] no needs to carry out calculations
in Cartesian coordinates so we can calculate in any coordinate system.

M\o ller's energy-momentum complex is given by

\begin{equation}
M_i^{\,\,\,k}={\frac 1{8\pi }}\cdot {\frac{\partial \chi _i^{\,\,\,kl}}{%
\partial x^l},}  \label{1.14}
\end{equation}
where 
\begin{equation}
\chi _i^{\,\,kl}=\sqrt{-g}\left( {\frac{\partial g_{in}}{\partial x^m}}-{%
\frac{\partial g_{im}}{\partial x^n}}\right) g^{km}g^{ln}.  \label{1.15}
\end{equation}
The energy in the M\o ller prescription has the expression

\begin{equation}
E=\int \hskip-7pt\int \hskip-7pt\int M_0^{\,\,\,0}dx^1dx^2dx^3={\frac 1{8\pi
}}\iiint {\frac{\partial \chi _0^{\,\,\,0l}}{\partial x^l}}dx^1dx^2dx^3.
\label{1.16}
\end{equation}

For the line-element given by (1) and (3) the $\chi _0^{\,\,\,01}$ component
is given by

\begin{equation}
\chi _0^{\;\,01}=\frac{(2Mr-2Mr\tanh (\frac{q^2}{2Mr})-q^2+q^2\tanh ^2(\frac{%
q^2}{2Mr}))\sin \theta }r.  \label{1.17}
\end{equation}

From (16) and (17) and applying the Gauss theorem we obtain the energy
distribution

\begin{equation}
E(r)=M(1-\tanh (\frac{q^2}{2Mr}))-\frac{q^2}{2r}(1-\tanh ^2(\frac{q^2}{2Mr}%
)).  \label{1.18}
\end{equation}

Also, from (18) we have

\begin{equation}
E(r)=M-\frac{q^2}r+\frac{q^6}{6M^2r^3}-\frac{q^{10}}{40M^4r^5}+O(\frac
1{r^6}).  \label{1.19}
\end{equation}

\section{Discussion}

Bondi [18] sustained that a nonlocalizable form of energy is not admissible
in relativity so its location can in principle be found. We obtain the
energy distribution of a charged regular black hole by using the
energy-momentum complexes of Einstein and M\o ller. From (13) it results
that in the Einstein prescription the first two terms in the expression of
the energy correspond to the Penrose quasi-local mass definition (evaluated
by Tod) [19] . The M\o ller formalism provides in the expression of the
energy (19) a term $M-\frac{q^2}r$ which agrees with the Komar [20]
prescription. Also, the M\o ller energy-momentum complex allows to make the
calculations in any particular system of coordinates. However, the energy of
the ABG black hole obtained by using the Einstein energy-momentum complex
contains the quantity $M-\frac{q^2}{2r}$ which is agreeing with linear
theory. Our result sustains the Virbhadra opinion [21] that ''the Einstein
energy-momentum complex is in a better bill of health'' and gives acceptable
results for many space-times. Also, Chang, Nester and Chen [22] showed that
the energy-momentum complexes are actually quasilocal and legitimate
expressions for the energy-momentum.

\textbf{Acknowledgments}

I am grateful to K. S. Virbhadra for his helpful suggestions.

\[
\]
\[
\]

\end{document}